# Influence of Light Soaking on Silicon Heterojunction Solar Cells with Various Architectures

Jean Cattin, Laurie-Lou Senaud, Jan Haschke, Bertrand Paviet-Salomon, Matthieu Despeisse, Christophe Ballif, Mathieu Boccard

*Keywords* — Silicon heterojunction, light soaking, reliability

*Abstract* — In this article we investigate the effect of prolonged light exposure on silicon heterojunction solar cells. We show that although light exposure systematically improves solar cell efficiency in the case of devices using intrinsic and p-type layers with optimal thickness, this treatment leads to performance degradation for devices with an insufficiently thick (p) layer on the light-incoming side. Our results indicate that this degradation is caused by a diminution of the (i/p)-layer stack hole-selectivity due to light exposure. Degradation is avoided when a sufficiently thick (p) layer is used, or when exposure of the (p) layer to UV light is avoided, as is the case of the rear-junction configuration, commonly used in industry. Additionally, applying a forward bias current or an infrared light exposure results in an efficiency increase for all investigated solar cells, independently of the (p) layer thickness, confirming the beneficial influence of recombination on the performance of silicon heterojunction solar cells.

## I. INTRODUCTION

Solar panels, once exposed to environmental conditions, must withstand several effects that may modify their performance over time. Therefore, understanding the mechanisms behind these effects is essential to prevent severe performance degradation. In the case of silicon heterojunction (SHJ) solar cells, intrinsic and doped hydrogenated amorphous silicon layers (a-Si:H) are used for surface passivation and carrier selectivity [1]. Such layers have been used in thin-film silicon solar cells and have exhibited performance degradation upon prolonged illumination [2], [3]. This degradation is due to a light-induced increase in defect density in a-Si:H, discovered in 1977 by Staebler and Wronski [4]. This effect is typically responsible for the ~10 % - 20 % relative conversion efficiency degradation observed in thin-film silicon modules during the first months of light exposure and is fully reversible upon annealing [5], [6]. In the case of intrinsic a-Si:H deposited on an n-type crystalline silicon wafer (c-Si($n$)), light exposure affects surface recombination of the charges photogenerated in the c-Si($n$) wafer, due to the light-induced creation of defects in the a-Si:H($i$) layer [7], [8]. Furthermore, differences in defect creation kinetics have been measured between the bulk a-Si:H layer and the interface defects, highlighting the significant contribution of interfaces when assessing light-induced modifications in SHJ cells [9].

In the case of poor initial passivation, the surface recombination rate has been observed to decrease after light soaking, revealing a decrease in the density of active interface defects, even though the density of bulk a-Si:H($i$) layer defects increased. This was tentatively attributed to higher hydrogen mobility in the layer under illumination, promoted by weaker Si-H bonds, due to trapping of photogenerated charges [10]. In turn, this causes hydrogen migration towards defect-rich regions until reaching equilibrium, changing the bulk/interface defect ratio and leading to a lifetime improvement, if the interface defect density was initially high. In the case of a well-passivated interface, the level of charge injection from c-Si into a-Si:H is higher, which induces more defects in the bulk a-Si:H layer than with a defective interface. This results in hydrogen diffusion from the interface to the bulk and, eventually, a lifetime reduction [9].

When a doped layer is present on top of the passivating a-Si:H($i$) layer, light soaking leads to a passivation improvement. Furthermore, the improvement rate is slower for thicker a-Si:H($i$) layers [9]. This effect was attributed to differences in the band structure induced by the doped layer and the subsequent electric field at the interface. The diffusion of atomic hydrogen in a-Si:H films was reported to be affected by the Fermi level position and to an eventual electric field, due to mobile $H^-$ [11]. It is still believed that the light soaking improvement is caused by an electric field, forcing hydrogen diffusion towards the interface, favoured by the larger hydrogen mobility induced by the trapping of light-generated charges and a subsequent reduction of the dangling bonds density [12]. Potentially, the passivation improvement could be explained by an increase in the charged defect density close to the interface, resulting in an improved so-called field-effect [13].

At the device level, industrial-grade cells have been reported to exhibit both increased open-circuit voltage ($V_{OC}$) and fill-factor (FF) during the first days of light soaking treatment, directly resulting from the aforementioned surface-passivation improvement [14], [15]. Interestingly, a similar increase could be reached without light, by applying a forward electric bias on the cell, with a current density equivalent to the short-circuit current density under AM1.5G spectrum illumination ($J_{SC}$). This similarity indicates that the passivation improvement is related to carrier recombination, injected from the c-Si wafer, rather than direct light illumination, similarly to bulk a-Si:H modification [16]. Tracking the evolution of minority carrier lifetime, $V_{OC}$ and *FF* as a function of light soaking time indicated a stabilisation after four days. Contrarily to light-induced degradation of a-Si:H thin-film solar cells, the light-induced improvement of SHJ devices is stable upon annealing.

Some studies additionally report on SHJ module degradation over time during outdoor operation [17]. The reported degradation remains within the module warranty and is similar to the values observed in other c-Si-based technologies. The degradation is faster during the first two years and mostly affects the module's $V_{OC}$ and series resistance ($R_S$). While some degradation mechanisms are related to the module metallisation and encapsulation, the substantial $V_{OC}$ drop is supposedly caused by passivation layer degradation. Such observations require more investigations to understand the possible degradation mechanisms in SHJ devices, both at the cell and module level.

Although the kinetics of device improvement during the first days are known, and good insights into the responsible mechanisms exist, the influence of different a-Si:H layer thicknesses remains to be determined, notably as it has been shown to play an important role in the case of thin-film a-Si:H devices: a $V_{OC}$ improvement could be achieved by light soaking devices with suboptimal *(p)* layer thickness [18]. This gain was attributed to a beneficial increase in dangling bond density in the *(p)* layer. Motivated by this observation, we focused on investigating the influence of the intrinsic *(i)* and *(p)* layer thicknesses on light-induced and forward bias changes in SHJ solar cells. In this study, we confirm the light-induced performance boost of well-optimised devices and highlight a previously unreported light-induced degradation of the hole-contact selectivity when thinner-than-optimal (p) layers are used.

## II. EXPERIMENTAL DETAILS

The solar cells presented in this study were all fabricated using float-zone, 180-µm-thick, n-type (2 Ω·cm) silicon wafers, textured by alkaline etching. Prior to a-Si:H layer deposition, an HF solution was used to remove native oxides from the surfaces. The a-Si:H layers were deposited in a plasma-enhanced chemical vapour deposition (PECVD) reactor. When mentioned, a p-type microcrystalline silicon layer (µc-Si:H(p)) is deposited instead of the a-Si:H(p) layer.

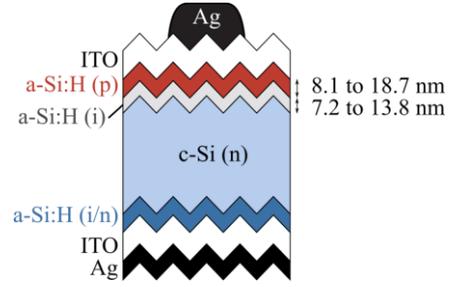

**Figure 1:** Schematic side view of a typical front-junction SHJ cell (not to scale).

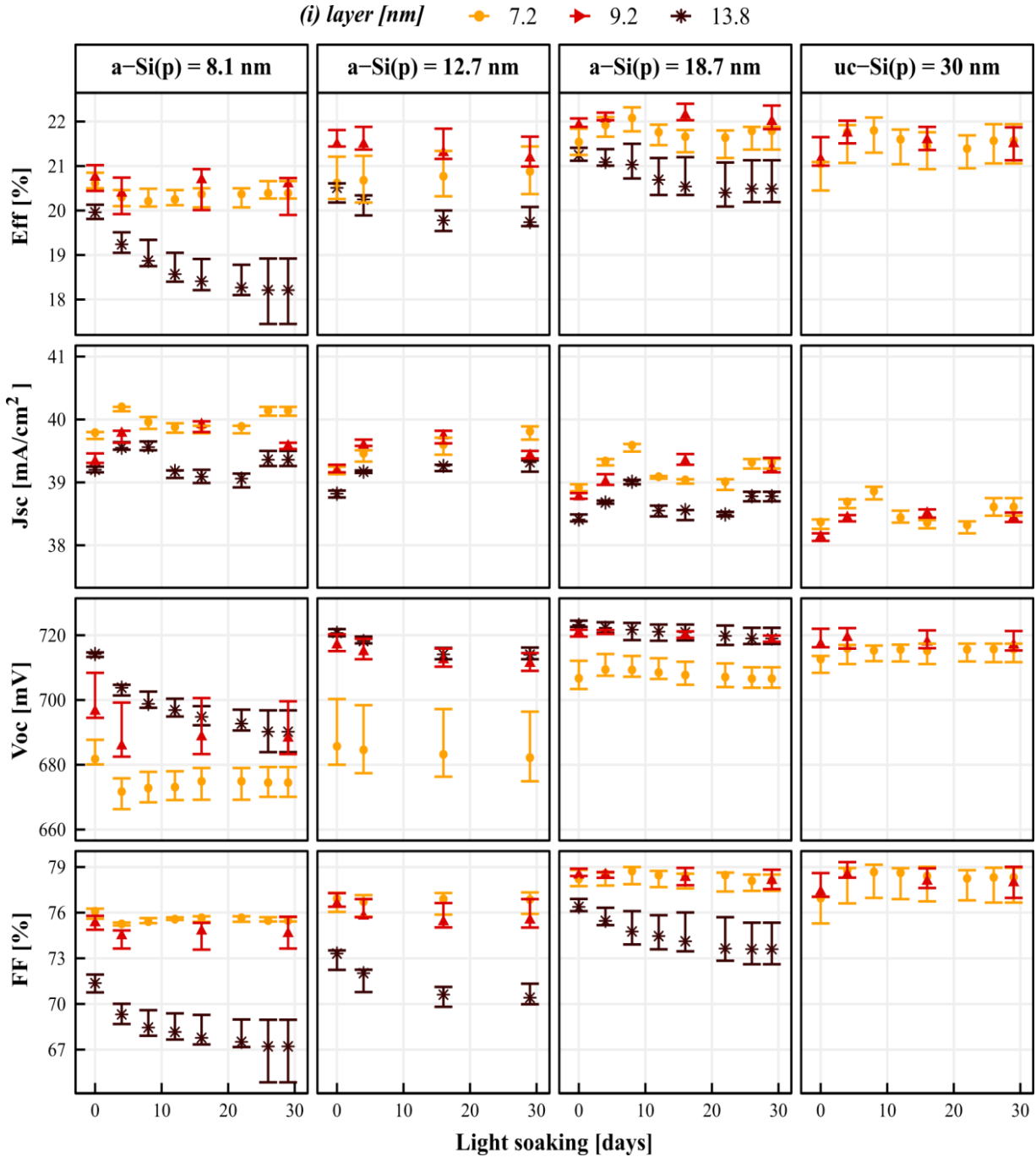

**Figure 2:** J-V parameters for solar cells employing different *(i)* and *(p)* layer thicknesses as a function of light soaking time. The dots represent the median, and the error bars the minimum and maximum values. The far-right values (µc-Si:H*(p)* of 30 nm) corresponds to samples with a microcrystalline silicon *(p)* layer instead of an amorphous silicon layer.

The rear side consists in a 9 nm (i) layer covered by a 30 nm (n) layer. Indium-tin oxide (ITO) layers and silver metallisation on the rear side were deposited via DC magnetron sputtering. A metallic shadow mask was used during deposition of the ITO layers to define five 4 cm$^2$ solar cells per wafer. When fabricated in rear-junction configuration, the rear ITO and silver are also masked following the cells pattern. The front metallisation was made of screen-printed silver paste cured at 210 °C for 30 minutes. The resulting device is schematically represented in Figure 1. The system used for the light soaking process is a *Solaronix Solixon A-65* degrader, delivering a class A AM1.5G spectrum at 1000 W/m$^2$, with a spatial nonuniformity of class B (IEC 60904-9). The sample temperature is regulated to 50 °C using an aluminium chuck with a water circulation system connected to a *Julabo FC600S* chiller, allowing for a temperature accuracy of ±0.2 °C. If not stated, the light soaking treatment is performed on cells under open-circuit (OC) conditions. When mentioned, cells are subject to a forward electric bias treatment, which consists of applying a forward current density of 40 mA/cm$^2$ (corresponding to the minority carrier injection level reached at $V_{OC}$ under an AM1.5G illumination) in the dark through aluminium foil contacts using *TPAE IC-triple* power supplies. Forward bias is applied at ambient temperature with minimal sample heating from dissipation of the applied electric power (~30 °C).

The $FF_0$ is calculated indirectly by performing multiple IV measurements at different irradiances, then comparing the $V_{OC}$ at high and low illumination, as described in [19], [20]. It is similar to the pseudo-FF (*pFF*) measured by suns-$V_{OC}$ and represents the *FF* of a cell without the resistive losses related to the $R_S$ [21]. The $R_S$ at the maximum power point (MPP) is extracted from the same analysis. When mentioned, dedicated lifetime samples were co-processed on quarter wafers with the cell precursors. These samples have no metallisation layer and were all annealed at 210°C for 30 minutes to match the post-metallisation process used on cells.

### III. RESULTS & DISCUSSION

*A) Effect of light soaking: Influence of layer thickness*

Eleven samples were produced in front-junction configuration. Nine samples used an a-Si:H(*p*) front contact with three different thicknesses for both the a-Si:H(*p*) and the front a-Si:H(*i*) layers, whereas two samples used a µc-Si:H(p) layer. The thicknesses are listed in Table 1:

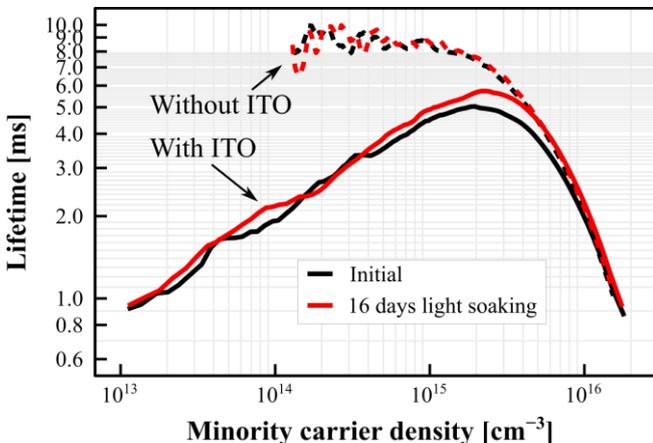

**Figure 3:** Lifetime curve measured on dedicated samples without silver metallisation, with and without ITO coverage. The samples have a thick (i) and a thin (p) layer, however, the other combinations (not shown here for simplicity) show similar behaviour.

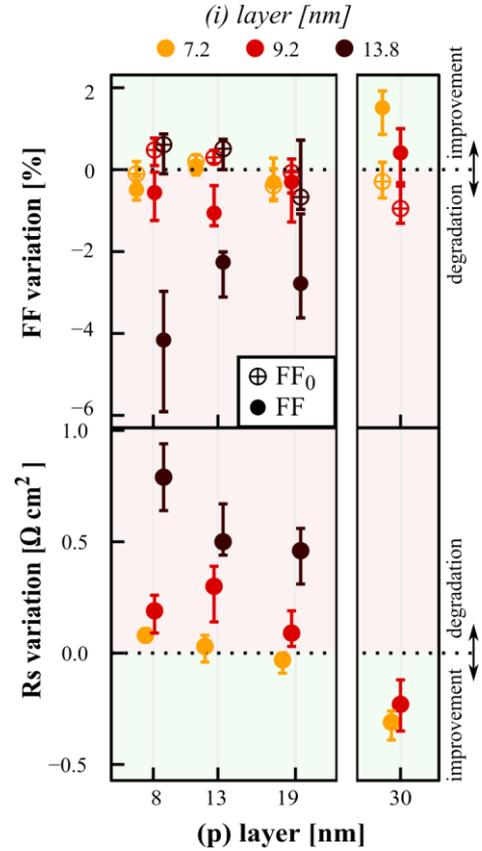

**Figure 4:** Absolute *FF*, $FF_0$ and $R_S$ variations caused by a light soaking treatment of four weeks. The dots and error bars represent the median and extremum values. A small shift was introduced in the x-axis values to differentiate the *(i)* layer thicknesses.

Table 1: Thickness variation of the a-Si:H(i,p) layers.

| layer | Thickness (nm) |
|---|---|
| **a-Si:H(i):** | 7.2, 9.2 and 13.8 |
| **a-Si:H(p):** | 8.1, 12.7 and 18.7 |
| **µc-Si:H(p)** | 30 nm |

Figure 2 shows the evolution of the J-V parameters as a function of the light soaking time. The a-Si:H(i,p) thickness ranges were chosen to be relevant to devices above 20% efficiency, and one of the conditions (9.2-nm-thick a-Si:H(i) with 18.7-nm-thick a-si:H(p)) exhibits >22% efficiency.

*1) Initial performance and general trends*

Looking at initial performances, observations similar to those previously reported in literature can be made: thickening the *(i)* layer leads to higher $V_{OC}$ due to improved surface passivation [22], but increases the $R_S$ [23]. A thicker *(p)* layer improves both $V_{OC}$ and *FF* through enhanced carrier selectivity [24], [25]. The $J_{sc}$ decreases linearly with the a-Si:H stack thickness, at a rate of -0.07 mA/cm$^2$ per nanometre of a-Si:H, the *(i)* and *(p)* layers thicknesses having the same impact on the measured $J_{sc}$. This result contrasts with earlier studies, which showed that a fraction of the carriers generated in the *(i)* layer could contribute to the cell current [26], [27]. Our observation either reflects the absence of charge collection from the *(i)* layer used in our study, or by a difference in absorption coefficient in the *(i)* and *(p)* layers compensating for the partial collection of charges generated in the *(i)* layer.

Looking at the evolution of *J*sc, although little spread is observed on a given day, trendless fluctuations are observed along time which are attributed to measurement uncertainty for days-apart measurements (±0.3%). These uncertainties prevent us to conclude definitively on an effect of the light soaking treatment, yet we can conclude that the possible changes are independent on the (i) or (p) layer thickness and below ±0.3%. Current fluctuations only have a minor impact on the efficiency variation and focus will be mainly placed on the $V_{OC}$ and *FF* in the following discussion.

The *(p)* layer thickness similarly modifies the light-induced changes of $V_{OC}$ and FF. Contrarily to previous studies, most samples degraded during light soaking. This degradation is most substantial for thin (p) and thick *(i)* layers (up to -23 mV $V_{OC}$ loss and -4 %$_{absolute}$ *FF* loss after four weeks of light soaking). On the one hand, these losses could be caused by the small hole reservoir and thick screening layer of these samples, as will be discussed later. On the other hand, samples with a thick *(p)* and thin *(i)* layer show *FF* and $V_{OC}$ improvement leading to a small gain in efficiency.

Figure 3 shows lifetime measurements performed on dedicated samples under light soaking, with and without ITO, for a specific architecture ("thick" *(i)* layer of 13.8 nm, "thin" *(p)* layer of 8.1 nm) known to show degrading J-V parameters under light soaking. After a 16 days light soaking treatment, the lifetime slightly improved or remained stable but was in no case reduced, indicating that surface passivation changes are not responsible for the observed $V_{OC}$ and *FF* losses.

For the *FF*, in addition to Figure 2, the variation caused by the light soaking treatment (final state minus initial state) is also shown in Figure 4, separated in $FF_0$ (related to passivation) and $R_S$ at MPP (related to charge transport) [20]. For all samples, congruently with lifetime measurements, $FF_0$ was not significantly modified by light soaking; most of the *FF* changes come from $R_S$ variation.

*2) Influence of the a-Si:H(i) layer thickness*

Figure 2 shows substantial degradation of samples with a thick (i) layer. The $V_{OC}$, however, remains higher with a thick rather than with a thin *(i)* layer, despite a more pronounced degradation, due to better initial passivation. A thicker *(i)* layer causes larger $R_S$, already in the initial state. The light soaking treatment results in a substantial $R_S$ increase, causing a FF drop as discussed in next section. Interestingly, the median (optimized for best performance) *(i)* layer of 9.2 nm provides similarly good initial passivation as the thicker *(i)* layer, however, still shows little degradation, as the thin *(i)* layer.

Samples with thin *(i)* layers are less sensitive to light soaking than the ones with thick *(i)* layers, and show no significant $FF_0$ or $R_S$ variations. We attribute this to a lower density of light-induced defects in thin *(i)* layers, which, assuming a similar *(p)* layer, experience a stronger electric field than thick *(i)* layers [9]. In a-Si:H, light-induced defects are typically dangling bonds in the bulk of the layer. It results in amphoteric defects in the electronic structure that are positively charged, as the Fermi level is shifted towards the valence band by the influence of the *(p)* layer [28], [29]. In theory, these fixed positive charges could partially screen the field induced by the fixed negative charges (ionised acceptors) in the *(p)* layer and thus result in a lower carrier selectivity.

Figure 2 additionally shows that the temporal evolution of the degradation depends on the *(i)* layer thickness. In regard to the $V_{OC}$ and *FF*, the samples with a thick *(i)* layer reach a steady state after four weeks, whereas the other samples stabilise after about four days. Furthermore, a drop after four days followed by slight and slower improvement is observed for $V_{OC}$ and *FF* of the sample with a thin *(p)* and thin *(i)* layer. These differences are currently not understood, but we suspect that the different phenomena taking place (detailed in part B) have different kinetics. Measurements with shorter time intervals are needed to understand these kinetic effects.

*3) Influence of the a-Si:H(p) layer thickness*

The aforementioned $R_S$ increase observed for samples with the thickest *(i)* layer is enhanced when thinning down the *(p)* layers. This behaviour could partly be caused by a decrease in conductivity of the *(i)* layer and the neighbouring interfaces upon light soaking [4]. This $R_S$ increase is correlated with a contact selectivity lowering as the same samples also experienced a substantial $V_{OC}$ degradation without lifetime drop, this aspect is developed further. Even in the initial state, the samples with a thin *(p)* layer have a low $V_{OC}$ and *FF*, as the thinnest *(p)* layer does not provide enough selectivity to allow the $V_{OC}$ to get close to the $iV_{OC}$ value (683 mV $V_{OC}$ in spite of a 703 mV $iV_{OC}$). In these selectivity-limited samples, even a minor selectivity change has a noticeable impact on the $V_{OC}$. Despite an $FF_0$ increase (assessed later), the FF is reduced in the same samples, possibly due to a decrease in the *(p)* layer effective doping. A depletion zone is present on both sides of the *(p)* layer, due to the ITO band mismatch and the c-Si wafer electric field. It was shown that if the *(p)* layer is insufficiently doped or too thin to screen the Schottky barrier, the *(p)* layer gets fully depleted, which decreases its effective doping. This decreases the contact selectivity and impacts both the $V_{OC}$ and *FF* [30]–[32].

The $R_S$ increase can tentatively be explained by a decrease in the Poole-Frenkel current, as the trap density increases in the a-Si:H layer during light soaking. The Poole-Frenkel current is, with multi-step tunnelling through defect states, a predominant charge-transport mechanism through the heterocontact. Interestingly, the Poole-Frenkel current is proportional to the local electric field [33], which may explain why the effect of light soaking is larger in devices with a thin *(p)* layer and, therefore, are expected to have a lower electric field at the interface.

Comparing these results on SHJ solar cells to literature results on thin-film a-Si:H devices, Stuckelberger et al. showed that the *(p)* layer thickness has an opposite influence on $V_{OC}$ upon light soaking [18]: A $V_{OC}$ increase was observed after light soaking on samples with an insufficiently thick (p) layer, whereas a $V_{OC}$ decrease was observed for optimal or thick *(p)* layer samples. In this case, the samples with a thick *(p)* layer performed better already in the initial state, as they were not limited by selectivity. Light-induced degradation comes from another phenomenon than in SHJ solar cells, namely absorber degradation. In the case of a thin *(p)* layer, the device is selectivity-limited, therefore less sensitive to bulk quality variations and more sensitive to selectivity variations. The selectivity increase in devices with a thin *(p)* layer arises from a transfer of negative charges from the μc-SiO$_x$:H(p) to the a-SiC$_x$(p) layer present in the doped layers stack. It results in a larger quasi-Fermi level splitting and, therefore, larger $V_{OC}$. Since the devices presented here have a single *(p)* layer, the common observation is that for thin-*(p)*-layer samples, the effect of light-soaking increase, and the layer adjacent to the (p)-layer plays a role in the modification of the device performance.

### 4) Microcrystalline (p) layer results

In the results presented in section 2 and 3, the main parameter affecting the influence of light soaking is the *(p)* layer thickness. Notably, small improvements were observed only on samples with sufficiently thick *(p)* layers. Comparison samples were produced with a 30 nm thick µc-Si:H*(p)* layer. Microcrystalline layers are known to have a better doping efficiency than amorphous layers [34]. A higher active doping density results in a larger charge reservoir, which results in a lower light soaking sensitivity using µc-Si:H(p) compared to a-Si:H(p) layer at equivalent thickness. The results corresponding to these samples are shown on the right side of Figure 2. These samples show a slight improvement in both the $V_{OC}$ and the *FF*, the latter coming from a reduction of $R_S$ (no $FF_0$ change was observed). The mechanism behind the reduction of $R_S$ is not yet identified.

Figure 5 summarizes these effects on efficiency. While adequately designed solar cells improve during light soaking (cells with initial efficiency ~22% improve slightly), poorly designed cells can either improve or degrade (light soaking

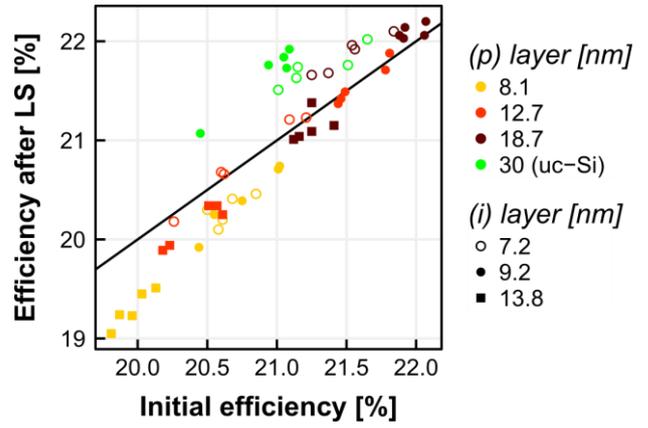

**Figure 5:** Final vs. initial absolute efficiency showing the effects of four weeks light soaking on different samples. The points above and below the line show improvement and degradation, respectively.

on 21%-initial-efficiency cells can result in anything between 20.7% and 21.9%). It also reveals a trade-off in the optimal *(p)* layer thickness between selectivity and parasitic light absorption. Thus, we observe that for thick microcrystalline *(p)* layers the final efficiency might be lower than cells with an optimised a-Si:H(p) layer, caused by large parasitic light absorption, despite the more important light soaking gain, as shown by the green dots in Figure 5. Note that this is a matter of optimization, since devices using recently optimized

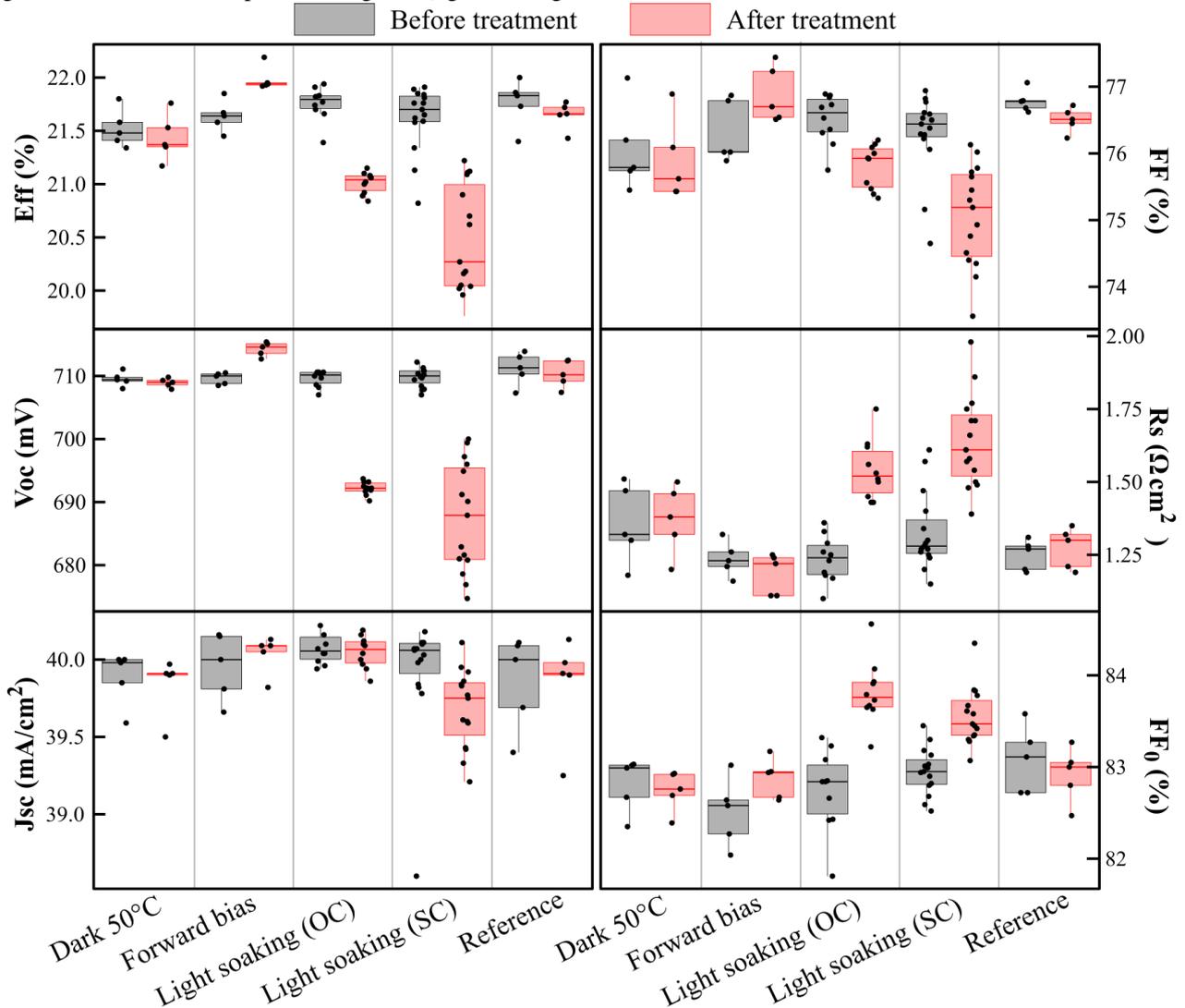

**Figure 6:** *J-V* results before and after different treatments on cells with a too-thin *(p)* layer.

ncSiOx:H(p) layers outperform devices using a-Si:H(p) layers while simultaneously improving under light soaking [35].

Our results suggest that to maximise the final efficiency and benefit from the light soaking treatment, the *(p)* layer should be slightly thicker than optimum, and the *(i)* layer as thin as possible. A *(p)* layer with a suboptimal thickness undergoes a $V_{OC}$ loss induced by a loss of selectivity and a FF loss due to an $R_S$ increase, which we try to assess in the following part.

### B) Decorrelation of light-induced degradation for SHJ with suboptimal a-Si:H(p) layer thickness

The standard light soaking process under open-circuit condition (later abbreviated as LS-OC) presented in the previous part is a combination of different physical effects listed hereafter. In this part, we decorrelate the individual effects and identify the mechanisms responsible for cell improvement and degradation. During a standard light soaking operation, cells are subjected to:

- **Air exposure:** cells are kept in ambient air for the duration of the experiment, without encapsulation.
- **Temperature:** cells are maintained at 50 °C in air.
- **Quasi-Fermi levels splitting** caused by charge generation in the bulk due to the one-sun illumination at open-circuit.
- **Light exposure:** the different layers of the device are exposed to light, with an impinging spectrum corresponding to AM1.5G.

The quasi-Fermi level splitting, due to the high-injection in the c-Si bulk, results in a high recombination rate and high density of trapped charges at the interfaces between c-Si and a-Si:H. This high injection state is known to be reproducible in the dark by exposing the cell to a forward electrical bias at a current density corresponding to the *J*sc [14]. The sole effect of light exposure has not yet been reported in literature. In an attempt to isolate the effect of light exposure on the cell without the contribution of a high-injection regime, some cells were connected in short-circuit while being exposed to light (later abbreviated as LS-SC). In this situation, photogenerated charges are extracted from the absorber, resulting in a steady-state minority-carrier density approximately three orders of magnitude lower than at open-circuit. While some residual minority carrier injection cannot be suppressed, this condition largely diminishes the effect of charges trapping and recombination at the interfaces, while maintaining the light exposure. Note that the cells were maintained at 50 °C during this treatment.

Figure 6 shows *J-V* parameters before and after seven days of different treatments. All the cells presented here were co-processed using a suboptimal *(p)* layer thickness of 7 nm, known to show a degradation of the J-V properties under LS-OC conditions, as discussed in the previous section. The (i) layer yields a thickness of 8 nm.

Several observations can be made for the samples with a suboptimal *(p)* layer thickness:

- The **initial $V_{OC}$ and FF** of all cells are low. This is caused by the suboptimal *(p)*-layer thickness, causing the cells to be selectivity limited with a low $iV_{OC}$ and $iFF$.
- The **reference samples** (kept in the dark at ambient temperature) and the samples exposed to **50 °C in the dark** are stable.

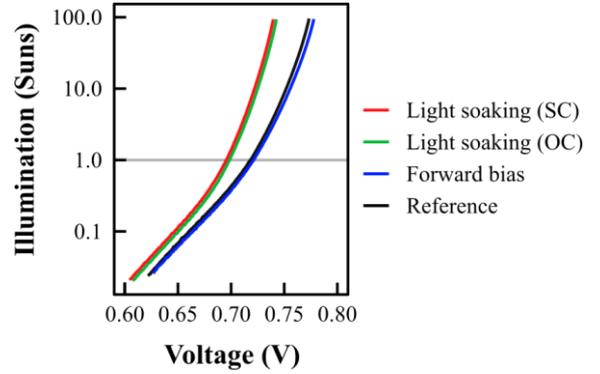

**Figure 7:** Suns-$V_{OC}$ measurements of cells with a too thin (p) layer after different treatments.

- All the cells exposed to **LS-OC** degrade accordingly to previous results.
- The cells exposed to **LS-SC** also degrade. Despite larger experimental noise than for LS-OC, the $V_{OC}$ and FF degradation are significantly larger (with statistical significance as shown with p-values of 5% and 0.4% respectively).
- The cells exposed to **forward bias** systematically show a *Voc* and *FF* improvement of 5.4 mV and 0.94 %$_{absolute}$, respectively.

These results suggest that two different mechanisms are responsible for cell improvement and degradation. The $R_S$ increase is similar for both LS conditions, and the $FF_0$ is increased for both conditions, especially for LS-OC. This surprising $FF_0$ increase is discussed further on in this work.

**Table 2:** Variations in the J-V results induced by the different treatments.

| Parameter | LS-SC | FW bias | LS-OC |
|---|---|---|---|
| $V_{OC}$ (mV) | -22.2±8.8 | +4.7±0.3 | -17.4±0.5 |
| FF (%$_{abs}$) | -1.22±0.58 | +0.57±0.09 | -0.71±0.14 |
| $R_S$ (Ωcm²) | +0.31±0.08 | -0.05±0.05 | +0.31±0.04 |
| $FF_0$ (%$_{abs}$) | +0.60±0.26 | +0.04±0.17 | +1.10±0.20 |

Table 2 summarises the differences induced by the different treatments on the *J-V* parameters. Interestingly, the effect of the LS-OC condition on the *J-V* results is closely reached by summing the effects of the LS-SC and forward bias conditions. This suggests that the impact of a standard light soaking treatment is the sum of two independent effects:

1) the high minority carrier injection in the bulk inducing recombination at the interfaces,
2) the exposure to light.

The high minority carrier injection positively affects the surface passivation, while the light exposure negatively affects the contact selectivity. Note that the standard deviation measured under each condition is too large to conclude on eventual temperature interactions, which we find here to have no measurable effect. Oliveau *et al.* [36] show that increasing the temperature (up to 120 °C) or the light intensity during an LS-OC treatment on optimised cells increases the lifetime improvement kinetics, thus showing that such interactions do exist at least for much higher temperatures than employed here.

The improvement measured after a forward bias treatment was reproduced with a similar magnitude on fully-optimised cells (with µc-Si:H(p) front-layer and double-layer anti-reflection coating) with an initial efficiency of 23.5%, suggesting that the improvements are independent of the initial efficiency on this efficiency range [35]. Additionally, the stability of the effects shown in figures 6 and 7 was studied by re-measuring the cells after four months in the dark at ambient temperature. These measurements show no significant modification compared to the post-treatment results. Therefore, the effects of light soaking (OC and SC) and forward bias can be considered as stable for the studied time scale and under room temperature condition.

*C) $FF_0$ increase under light soaking*

In this paragraph, we describe the $FF_0$ improvement measured after light soaking on cells with a suboptimal (p) layer which showed a decrease in Voc and FF due to a selectivity loss. FF0 is related to the ratio between the pseudo voltage at MPP (p$V_{MPP}$, calculated using a low-illumination J-V curve) and Voc. An increase in $FF_0$ is observed in our case despite a decrease in $V_{oc}$, FF and p$V_{MPP}$. This is simply because the decrease in pVmpp is smaller than the decrease of Voc, as confirmed by Suns-$V_{OC}$ measurements shown in Figure 7 for similar cells as the ones analysed above (with a too-thin (p) layer). It shows the $Voc$ as a function of illumination after different treatments. The $V_{OC}$ is reduced at all illuminations by both light soaking treatments. Furthermore, both light soaking treatments affect more the Voc at high illumination than at low illumination, which leads to the aforementioned $FF_0$ increase. This deviation from the logarithmic trend is well described by Bivour *et al.,* who show similar behaviour for cells with lowly doped *(p)* layers [25]. They attribute this phenomenon to a broader space charge region in the *(p)* layer on the ITO side, resulting in a reduction in contact selectivity and a larger recombination current within the *(i)* and *(p)* layers stack. If the selectivity loss is large enough, it may affect both the $V_{OC}$ and MPP under one sun, as observed in our samples. These results corroborate the conclusions of the first part of this work, which show that the light-induced defects reduce the selectivity of the (p) contact, resulting in a $V_{OC}$ and $FF$ loss. In Figure 7, we observe that there is no slope difference between the suns-$V_{OC}$ measurements of samples exposed to light soaking at open-circuit and short-circuit. A small shift of the suns-$V_{OC}$ towards higher voltages is measured on samples exposed to LS-OC compared to LS-SC. Similarly, the cell exposed to forward bias shows a small positive shift of the suns-$V_{OC}$ voltage compared to the reference cell, with no large difference in the suns-$V_{OC}$ slope. This is a signature of the improvement caused by high minority carrier injection and confirms that the selectivity loss is solely induced by the light exposure, as demonstrated in the next section.

*D) (p) layer exposition to light*

Here, we show the effect of placing the (p) layer at the front- or rear-side of the device, as well as exposing samples with a too thin (p) layer to UV and IR light. A new set of samples was produced with thin (p) layer (7 nm) and standard (20 nm) (p) layers, in rear and front emitter configuration. The (n) layer is similar as used in the previously shown samples. The J-V properties were measured before and after a one week treatment under various light sources (described below) in open-circuit condition.

The UV and IR sources were made of LED emitters at 365 nm and 860 nm, respectively. The intensity of the UV source was set to match the number of photons absorbed in the front layers (ITO and a-Si:H) under an AM1.5G spectrum. The IR source's intensity was set to reach the cell's Voc under standard test conditions, implying a similar minority carrier injection in the absorber. The standard light soaking is referred to as LS.

Figure 8 shows the absolute differences induced on the Voc and FF by the treatments on the different samples. The observations on front-junction samples under LS corroborate the conclusions from the previous sections. However, rear-junction samples all show small improvements with no measurable influence of the (p) layer thickness. This result indicates that protecting the (p) layer from direct exposition prevents the aforementioned selectivity loss of the (p) contact. It suggests that this contact degradation is not caused by high minority carrier injection in the absorber (similar in front and rear junction), but rather from direct light absorption into the a-Si:H layers on the (p) side. Note that this performance degradation upon illuminating the (p) layer was also measured on p-type wafers (data not shown here), for which the (p) layer is not forming the p-n junction, indicating that this is due to intrinsic properties of the (p) layer and not to the fact that this layer is the so-called "emitter" (i.e. the layer of opposite polarity compared to the wafer).

The samples exposed to UV light show a large difference between the front- and rear-junction configurations. While the front-junction samples degrade significantly, rear-junction ones remain unaffected by the treatment. As the UV light maximises light absorption into the front layers, while minimising quasi-Fermi levels splitting in the absorber, it isolates the effect of direct light absorption in the front layers, similarly to the LS-SC condition in section B. In rear-junction configuration, it can reasonably be assumed that the (p) layer is exposed to a negligible amount of light, which explains the insensitivity to the treatment.

The front-junction cells exposed to UV exhibit significantly higher degradation than front-junction cells exposed to LS. The small passivation benefit from LS is not enough to fully explain this difference. The total number of photons absorbed in the front layers is similar in both cases, however, in the LS case a significant share of these photons are at higher wavelengths (i.e. lower energies) than 365 nm. A higher energy is therefore absorbed in the front a-Si layers in the UV case than in the LS case, which could explain this larger performance degradation.

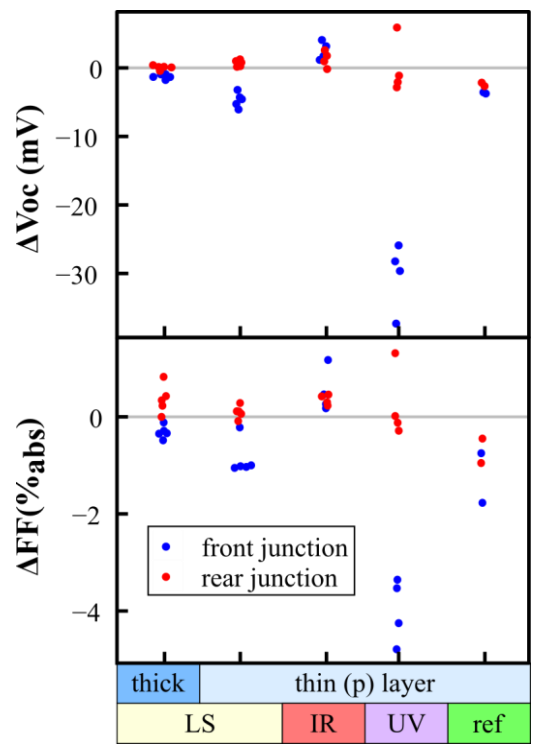

**Figure 8:** Absolute variation of the J-V parameters caused by different treatments on cells with a too-thin (p) layer.

The samples exposed to IR light all show little improvement with no significant influence of the junction side, similarly to what was observed on the samples under forward bias in the previous sections. At these wavelengths, the absorption coefficient in the a-Si:H layers is small, while still high in the c-Si. Therefore, similarly to the forward bias treatment, exposure to IR light induces splitting of the quasi-Fermi levels in the c-Si, while maintaining a negligible direct light absorption in the a-Si:H layers.

We demonstrate here that rear-emitter configuration effectively prevents potential degradation of poorly optimised (p) contact stacks, even in the case of bifacial cells, as the spectrum of light reflected back from the ground is UV-poor compared to AM1.5G [37], [38]. However, further experiments would be needed to validate this point.

Moreover, the nature of the detrimental defects observed when exposing the (p) layer side to UV light is still unclear. Dedicated characterisations of the material properties of a-Si(i) and a-Si(p) layers within a solar cell structure would be needed to identify the modifications induced by UV light in this context.

## IV. CONCLUSION

In this work, we showed that, for well-optimised silicon heterojunction solar cells, light soaking treatment results in improved efficiency. However, this treatment can also cause degradation, e.g. when the *(p)* layer is too thin and exposed to light. The standard light soaking environment can be split into two main components: light exposure and minority carrier injection. We decorrelated the effects of these two mechanisms and showed that they are independent from each other. In open-circuit, all the light-generated charges recombine in the device, mostly at the interfaces and in the contacts. Recombination of charges injected in the c-Si wafer, either through light-soaking, forward bias or IR light exposure, has a systematically positive impact on the cells, even in the case of non-optimal layers thickness. The light exposure of the *(p)* layer (isolated by performing a light soaking treatment in short-circuit conditions or UV light exposure) causes a selectivity reduction, which we tentatively attribute to defects generated by UV absorption in the *(p)* layer or *(i)* layer on the junction side, counteracting the active doping. This selectivity loss affects both the $V_{OC}$ and the *FF* of the cells using an insufficiently thick *(p)* layer in front-junction configuration. The cells with sufficiently high selectivity are not affected. We show that rear-junction cells are not affected, even with a poorly optimised *(p)* layer, as the *(p)* contact is protected from UV light exposure. According to our results, well-designed SHJ-based PV modules are expected to show a $0.3\%_{absolute}$ efficiency increase during the first week of operation. This positive effect can also be obtained by a post-production light exposure or forward electrical bias treatment of four days.

## ACKNOWLEDGEMENTS

The authors acknowledge Nicolas Badel, Patrick Wyss and Christophe Allebé for the high-quality wet-processing and metallisation, and Vincent Paratte for sample preparation help. We also thank Cédric Bucher and Aymeric Schafflützel for technical support, and Caroline Hain for gracefully reviewing the English.

This work was funded by Qatar Foundation, the European Union's Horizon 2020 research and innovation programs under Grant Agreement No. 745601 (Ampere) and by Swiss national science foundation under Ambizione Energy grant ICONS (PZENP2_173627).